\documentclass[aps,groupedaddress,preprint,superscriptaddress,showpacs]{revtex4-1}
\usepackage{eurosym}
\usepackage{amsmath}
\usepackage{fixmath}
\usepackage{graphicx}
\usepackage{wrapfig}
\usepackage[toc,page]{appendix}
\usepackage{subcaption}

\graphicspath{{images/}}

\newcommand{\be}{\begin{equation}}
\newcommand{\ee}{\end{equation}}

\begin{document}

\title{Four-wave mixing in Weyl semimetals}

\author{Sultan Almutairi}
\affiliation{Department of Physics and Astronomy, Texas A\&M University, College Station,
TX, 77843 USA}
\author{Qianfan Chen}
\affiliation{Department of Physics and Astronomy, Texas A\&M University, College Station,
TX, 77843 USA}
\author{ Mikhail Tokman}
\affiliation{Institute of Applied Physics, Russian Academy of Sciences, Nizhny Novgorod, 603950, Russia }
\author{Alexey Belyanin}
\affiliation{Department of Physics and Astronomy, Texas A\&M University, College Station,
TX, 77843 USA}

\begin{abstract}
Weyl semimetals (WSMs) have unusual optical response originated from unique topological properties of their bulk and surface electron states. Their third-order optical nonlinearity is expected to be very strong, especially at long wavelengths, due to linear dispersion and high Fermi velocity of three-dimensional Weyl fermions. Here we derive the third-order nonlinear optical
conductivity of WSMs in the long-wavelength limit and calculate the intensity of the nonlinear four-wave mixing signal as it is transmitted through the WSM film or propagates away from the surface of the material in the reflection geometry. All results are analytic and show the scaling of the signal intensity with variation of all relevant parameters. The nonlinear generation efficiency turns out to be surprisingly high for a lossy material, of the order of  several mW per W$^3$ of the incident pump power. Optimal conditions for maximizing the nonlinear signal are realized in the vicinity of bulk plasma resonance. This indicates that ultrathin WSM films of the order of skin depth in thickness could find applications in compact optoelectronic devices. 
\end{abstract}

\date{\today }
\maketitle

\section{Introduction}

Weyl semimetals (WSMs) are fascinating new materials with nontrivial topology of both bulk and surface electron states \cite{Wan2011, Burkov2011, Xu2015, Lv2015, Yan2017, Hasan2017, Armitage2018, Burkov2018}. Although most of the research on WSMs has been focused on their electronic
structure and transport, a number of recent studies have suggested that WSMs should also have highly unusual optical properties; see e.g.~\cite{kargarian2015, hofmann2016, tabert2016-2, ukhtary2017, felser2017, kotov2018, andolina2018,zyuzin2018, rostami2018, chen2019, narang2019,ma2019,moore2019,chen2019-2} and references therein. Their optical response can be used to provide detailed spectroscopic  information about their electronic structure which is in a sense complementary to the one obtained from transport studies. Furthermore, strong anisotropy and gyrotropy of their optical response in combination with strong optical nonlineairites \cite{chen2019, chen2019-2,tokman2020} makes WSM films promising for applications in optoelectronics and quantum technologies. 

Due to the presence of one or more pairs of separated Weyl nodes, the electron bandstructure of WSMs is anisotropic and includes both bulk and surface states. As a result, even in the weak-field linear regime the optical conductivity tensor is quite complicated and generally cannot be expressed analytically even within the simple microscopic model of a two-band  WSM  Hamiltonian with two separated Weyl nodes (\cite{Burkov2011,okugawa2014}); see e.g.~\cite{chen2019,chen2019-2} where bulk and surface conductivity tensors were derived and the properties of bulk and surface electromagnetic eigenmodes were described. Fortunately, one expects the strongest nonlinear optical response in the high-doping, long-wavelength limit $\hbar \omega \ll 2E_F, \hbar v_F b$, where $E_F$ is the Fermi energy and $2\hbar b$ is the Weyl point separation in momentum space \cite{chen2019}. In this limit the electron bandstructure takes a simple universal form of 3D cones for both Dirac and Type I Weyl semimetals. As a result, one can obtain analytic, although a bit cumbersome, expressions for the nonlinear conductivity of any order. 

There are two reasons why the nonlinear response is maximized in the long-wavelength limit. First, the magnitude of the dipole matrix element of the optical transitions scales as $\mu \sim e v_F/\omega$, where $v_F$ is the Fermi velocity and $-e$ is the  electron charge. This is true for both intraband and interband transitions. Note the linear scaling $\mu \propto \lambda$ for Weyl fermions as compared to the usual $\mu \propto \sqrt{\lambda}$ scaling for massive electrons with parabolic dispersion. As a result, the nonlinear $n$th order conductivity grows very rapidly, $|\sigma^{(n)}| \propto \mu^{n+1}$, with increasing wavelength. Taking into account the density of states, one can immediately predict the scaling $|\sigma^{(3)}| \propto \displaystyle \frac{e^4v_F}{(\hbar \omega)^3}$, which is confirmed below.  The resulting magnitude of $|\chi^{(3)}|$ is many orders of magnitude higher than in  conventional nonlinear materials.

The second reason is that at low frequencies $\hbar \omega \ll 2E_F$ the interband absorption is eliminated by Pauli blocking. Furthermore, the electric field of the nonlinear signal is enhanced in the vicinity of bulk plasma resonance due to the boundary conditions at the interface. The latter effect is similar to the observed enhancement of Kerr index modulation and third-harmonic generation in so-called epsilon-near-zero materials; see e.g.~\cite{capretti2015,alam2016}.  An exceptionally high value of $|\chi^{(3)}|$ in combination with field enhancement at plasma resonance lead to a surprisingly high efficiency of the nonlinear generation, of the order of  several mW per W$^3$ of incident pump power. 

The third order conductivity has been calculated in \cite{Sun 2018} in the hydrodynamic limit and for frequencies lower than the scattering rate $1/\tau$. Here we are interested in the frequencies higher than the scattering rate, but still low enough to limit the response to the vicinity of the Weyl points, as argued above. Therefore, we need to use the kinetic approach. There is some controversy surrounding the kinetic theory of the third-order response. In \cite{Zhong 2019} the third-order conductivity in the terahertz spectral range was calculated for degenerate FWM ($\omega + \omega - \omega$) and third-harmonic generation processes. However, in a very recent paper \cite{zuber2020} the third-order conductivity was found to be zero in the low-frequency limit and zero result was rationalized by symmetry arguments. While the inversion symmetry prohibits the nonzero {\it second-order} response in electric-dipole approximation, we don't see any symmetry arguments that would require the third-order response of WSMs to be zero, even assuming perfectly isotropic conical dispersion near every Weyl point.  And indeed, we present a very general kinetic equation-based derivation of the third-order conductivity to show that it remains finite and in fact quite large in magnitude at low frequencies. 

In Sec.~II of this paper, we derive the general expression for the  third-order nonlinear conductivity by using the kinetic equation formalism for frequencies higher than the phenomenological relaxation rate. We then proceed in Sec.~III to calculate the four-wave mixing (FWM) signal power transmitted through a WSM slab or propagating away from the surface of the material opposite to the direction of incident pump beams  as a function of relevant parameters.  The Appendix
contains some details of the derivation of the third-order susceptibility. 

\section{Third-order nonlinear optical conductivity}

We consider the optical response of a doped WSM at frequencies $ 1/\tau < \hbar \omega < 2E_F$  that are low enough so that the electron excitations in the vicinity of each Weyl point satisfy the linear dispersion:
\begin{equation}
E_{s}=s\mathbf{v\cdot p},
\end{equation}%
where $s=\pm 1$ is for the conduction and valence bands, respectively. We assume for simplicity that the velocity has the same magnitude in every direction, i.e. the cone is isotropic. Anisotropic cones can be easily incorporated into the analytic theory below, but they will make the expressions more cumbersome without changing the nonlinear response qualitatively. We will assume for definiteness that the Fermi level is in the conduction band.   Thus we
have
\begin{equation}
\mathbf{v}=\frac{\partial E_{+}}{\partial \mathbf{p}}=v_{F}\mathbf{n},
\end{equation}
where $E_{+}$ is the electron energy in the conduction band, $p=\sqrt{p_{x}^{2}+p_{y}^{2}+p_{z}^{2}}$ is the magnitude of electron momentum, $v_{F}$ is the Fermi velocity, and $\mathbf{n}=\frac{
\mathbf{p}}{p}=(\sin \theta \cos \phi ,\sin \theta \sin \phi ,\cos \theta )$
is the unit vector in the direction of the electron velocity in spherical coordinates. For $\hbar \omega < 2E_F$ and in the limit of strong Fermi degeneracy, intraband transitions make the dominant contribution. When only the intraband transitions are included, the fully quantum approach based on the von Neumann equation for the density matrix gives the same result as the semiclassical kinetic equation approach. For massless 2D Dirac fermions this was checked explicitly in \cite{wang2016}.  The kinetic equation with a phenomenological collision term has a standard form: 
\begin{equation}
\label{kin} 
\frac{\partial f}{\partial t}+v_{F}(\mathbf{n\cdot \nabla })f-e[\mathbf{E}+%
\frac{v_{F}}{c}(\mathbf{n}\times \mathbf{B})]\cdot \frac{\partial f}{%
\partial \mathbf{p}}=\gamma \lbrack F(p)-f],
\end{equation}%
where $\mathbf{E}$\textbf{\ }and $\mathbf{B}$ are external electric and
magnetic fields, respectively, $\gamma $ is the electron relaxation rate, $F(p)$ is an unperturbed (zeroth-order)
distribution function, which is chosen as the equilibrium Fermi-Dirac
distribution, and $f$ is the non-equilibrium distribution function in the presence of external fields. The current density can be then calculated as
\begin{equation} \label{cur}
 \mathbf{j}(\mathbf{r},t) = -e \int \mathbf{v} f(\mathbf{r},\mathbf{p},t)\, d^3p. 
\end{equation}
We are interested in the electric-dipole optical response, so  the magnetic field term 
can be neglected. We also assume that the electric field has the form
\begin{equation}
\label{e}
\mathbf{E}(\mathbf{r},t)=\sum_{n}\mathbf{E}_{n}(\mathbf{r},\omega _{n})e^{-i\omega
_{n}t}=\sum_{n}\mathbf{A}_{n}e^{ik_{n}z-i\omega _{n}t},
\end{equation}%
and make an ansatz for the non-equilibrium distribution function: 
\begin{equation}
\label{f}
f=\sum_{m}\xi _{m}e^{iq_{m}z-i\omega _{m}t},
\end{equation}%
where we have set $\xi _{0}=F(p)$, $E_{0}=0$, $\omega _{0} = q_0=0$. Because both the
electric field and the non-equilibrium distribution function are real, i.e. $%
\mathbf{E}(\mathbf{r},t)=\mathbf{E}^{\ast }(\mathbf{r},t)$ and $%
f=f^{\ast },$ we obtain $E_{-n}=E_{n}^{\ast },$ $\xi _{-n}=\xi _{n}^{\ast },$
$\omega _{-n}=-\omega _{n}$, $q _{-n}=-q _{n}$.

Substituting Eqs.~(\ref{e}) and (\ref{f}) into Eq.~(\ref{kin}) and transforming into spherical coordinates, one can write Eq.~(\ref{kin}) in the following form: 
\begin{equation}
\xi _{n}=\sum_{m,k}G^{n,m}\xi _{k}.
\end{equation}
Here the operator $G^{n,m}$ is determined by
\begin{eqnarray}
G^{n,m}(p,\phi ,\theta ) &\equiv &g_{1}^{n,m}(\phi ,\theta )\frac{%
\partial }{\partial p}+g_{2}^{n,m}(\phi ,\theta )\frac{\partial }{%
p\partial \phi }  \notag \\
&&+g_{3}^{n,m}(\phi ,\theta )\frac{\partial }{p\partial \theta },
\end{eqnarray}
where $p,\phi ,\theta$ are spherical coordinates in momentum space and 
\begin{equation}
\label{g1}
g_{1}^{n,m}\equiv e\frac{E_{m,x}\cos \phi \sin \theta +E_{m,y}\sin \phi
\sin \theta +E_{m,z}\cos \theta }{-i\omega _{n}+\gamma },
\end{equation}
\begin{equation}
\label{g2}
g_{2}^{n,m}(\phi ,\theta )\equiv e\frac{E_{m,y}\cos \phi -E_{m,x}\sin \phi
}{\sin \theta (-i\omega _{n}+\gamma )}\frac{\partial }{p\partial \phi },
\end{equation}
\begin{eqnarray}
g_{3}^{n,m}(\phi ,\theta ) &\equiv &\left( E_{m,x}\cos \phi \cos \theta
+E_{m,y}\sin \phi \cos \theta -E_{m,z}\sin \theta \right)  \notag \\
&&\times \frac{e}{\left( -i\omega _{n}+\gamma \right) }\frac{\partial }{%
p\partial \theta }.
\label{g3}
\end{eqnarray}
Note that in electric dipole approximation we neglect the magnetic-field dependent terms and the terms with spatial gradients in Eq.~(\ref{kin}) which give rise to the Doppler shift in resonant denominators in Eqs.~(\ref{g1})-(\ref{g3}). Keeping such terms would lead to corrections that scale as powers of the small parameter $v_F/c$ \cite{wang2016} . 

The optical response in any order for an arbitrary non-degenerate multi-wave mixing can be calculated by the repetitive applying of  $G^{n,m}$ to the equilibrium distribution function. 
For example, the  first-order
approximation describing the linear optical response is 
\begin{equation}
\xi _{n}^{(1)}=G^{n,n}\xi _{0}=g_{1}^{(n,n)}\frac{\partial F}{\partial p}.
\end{equation}
Substituting this into  Eq.~(\ref{cur}) and using $\int_{0}^{\infty }\frac{\partial F(p)}{\partial p}
p^{2}dp=-p_{F}^{2}$ in the strong degeneracy/low temperature limit, one can get
\begin{equation}
\label{sigma1} 
\sigma ^{(1)}(\omega )=\frac{e^{2}v_{F}p_{F}^{2}g_{s}g_{w}}{6 \pi^2
\hbar^{3}(\gamma -i\omega )}, 
\end{equation}
where $g_{s}$ and $g_{w}$ are the degeneracy factors associated with spin and the number of Weyl nodes respectively. 

The second-order
approximation of the non-equilibrium distribution function is $\xi _{l}^{\left(
2\right) }$,
\begin{equation}
\xi _{l}^{(2)}=\sum_{m,k}G^{l,m}\xi _{k}^{\left( 1\right)
}=\sum_{m,k}G^{l,m}G^{k,k}\xi _{0}
\end{equation}%
for all possible $\omega _{m}$ and $\omega _{k}$ satisfying the relation $
\omega _{l}=\omega _{m}+\omega _{k}$.  Similarly, 
the third-order response is described by 
\begin{equation}
\label{third}
\xi _{i}^{(3)}=\sum_{j,m,k}G^{i,j}\xi _{l}^{\left( 2\right)
}=\sum_{j,m,k}G^{i,j}G^{l,m}G^{k,k}\xi _{0}
\end{equation}
for all possible $\omega _{j},\omega _{m}$ and $\omega _{k}$ satisfying the
relation $\omega _{i}=\omega _{j}+\omega _{m}+\omega _{k}.$

The
nonlinear current $\mathbf{j}(\omega _{n}=\omega _{1}+\omega _{2}+\omega
_{3})$ is then given by
\begin{eqnarray}
\left(
\begin{array}{c}
j_{x} \\
j_{y} \\
j_{z}
\end{array}
\right) &=&-ev_{F}\int_{0}^{\infty }\int_{0}^{2\pi }\int_{0}^{\pi }\xi
_{n}^{\left( 3\right) }\left(
\begin{array}{c}
\cos \phi \sin \theta \\
\sin \phi \sin \theta \\
\cos \theta%
\end{array}%
\right) p^{2} \sin \theta dp d\theta d\varphi.   
\end{eqnarray}
The integral is evaluating in the Appendix. The resulting third-order nonlinear optical
conductivity tensor has the form
\begin{eqnarray}
\sigma _{ijkl} &=& \frac{e^{4}v_{F}g_{s}g_{w}\Delta _{ijkl} 
}{90 \pi^2 \hbar^{3}(\gamma -i\omega _{1})}   \frac{1}{\left[ \gamma -i(\omega _{1}+\omega _{2})\right] \left[
\gamma -i(\omega _{1}+\omega _{2}+\omega _{3})\right] }  \nonumber \\ 
&+&  {\rm \, all \, permutations\, of\, } \omega_{1},\omega_{2},\omega_{3},
\end{eqnarray}
where $\Delta
_{ijkl}\equiv \delta _{ij}\delta _{kl}+\delta _{ik}\delta _{jl}+\delta
_{il}\delta _{jk}$. Here $\delta _{ij}$ is the Kronecker delta. 

In the particular case of the third harmonic generation $\omega _{1}=\omega _{2}=\omega
_{3}=\omega $. Then the nonlinear current at $\omega _{n}= 3\omega $ is
\begin{equation}
j_k^{(3)}(3\omega) =\frac{e^{4}v_{F}g_{s}g_{w}(E_{1,x}^{2}+E_{1,y}^{2}+E_{1,z}^{2})%
}{5 \pi^2 \hbar^{3} (\gamma - i3\omega )(\gamma - i2 \omega )(\gamma - i\omega )}%
E_{1,k},
\end{equation}
where $k = (x,y,z)$. This is consistent with the result for $\sigma _{3}^{intra}\left(
3\omega \right) $ in \cite{Zhong 2019} when $E_{1,x,y,z}=E_{0}.$

In another special case of partially degenerate FWM we consider the nonlinear current at frequency $\omega _{s}=2\omega _{1}-\omega _{2}$. For simplicity, we assume that the electric field is
along the $z$-axis; then the $z$-component of the nonlinear current is
\begin{eqnarray} 
j_{z}^{(3)}(\omega_s)
&=& \frac{e^{4}v_{F}g_{s}g_{w}E_{1,z}^{2}E_{2,z}^{\ast }}{%
15 \pi^2 \hbar^{3}(-i\omega _{s}+\gamma )}  \left[ \frac{\frac{1}{i\omega
_{2}+\gamma }+\frac{1}{-i\omega _{1}+\gamma }}{(-i(\omega _{1}-\omega
_{2})+\gamma )}  \right. \nonumber \\
&& \left. + \frac{1}{(-i2\omega _{1}+\gamma )(-i\omega _{1}+\gamma )} \right]. 
\label{sigma3}
\end{eqnarray}%

\begin{figure}[htb]
\begin{center}
\includegraphics[scale=0.5]{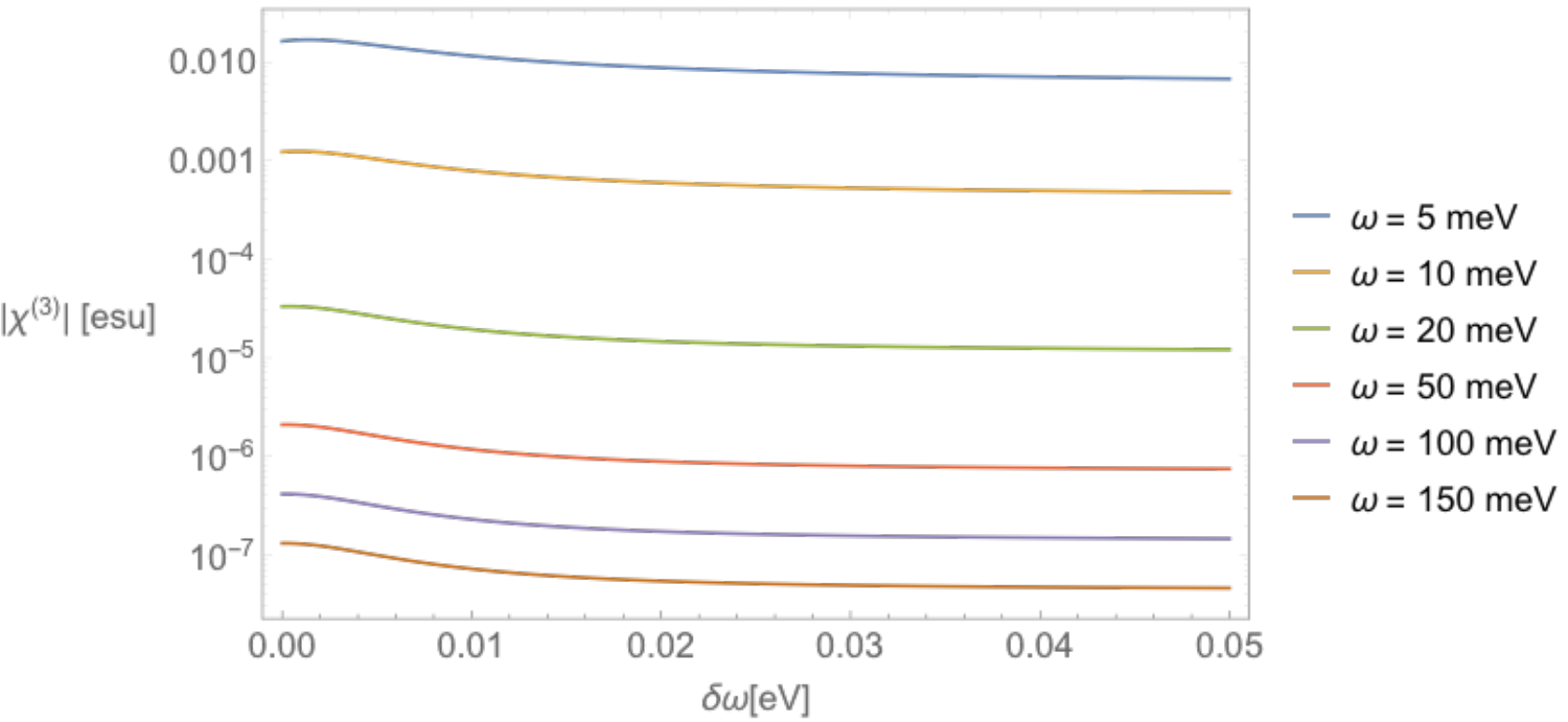}
\caption{The absolute value of $\chi^{(3)}$ as a function of detuning $\delta \omega = \omega_2 - \omega_1$ for several values of $\omega_1$. Other parameters are $\hbar \gamma =  5$ meV, $v_F = 10^8$ cm/s, $g_s = 2$, $g_w = 4$. }
\label{fig1}
\end{center}
\end{figure}

Note the resonance at  $\omega _{1} = \omega _{2}$.  The absolute value of the third-order susceptibility $\chi ^{(3)}=
\frac{i\sigma ^{\left( 3\right) }}{\omega_s }$ which follows from Eq.~(\ref{sigma3}) is plotted in Fig.~\ref{fig1} as a function of detuning $\delta \omega = \omega_2 - \omega_1$ for several values of $\omega_1$. The magnitudes of $\chi^{(3)}$ are many orders of magnitude higher as compared to typical values in the conventional nonlinear crystals \cite{Boyd2003}. Moreover, the numerical results in Fig.~\ref{fig1} and other figures below were obtained for $g_w = 4$, i.e. two pairs of Weyl points. Even higher nonlinearity and the nonlinear signal intensity are expected for larger values of $g_w$.  However, strong optical absorption in WSMs limits the nonlinear signal power, as we show in the next section. 

\section{Intensity and power of the four-wave mixing signal}

As the simplest problem relevant to the experiment, we consider two monochromatic pump fields at frequencies $\omega_1$ and $\omega_2$ normally incident at the WSM layer from the air. The case of an oblique incidence can be easily solved in the same way, but we will try to keep the expressions less cumbersome. The nonlinear FWM signal at frequency $\omega_s = 2 \omega_1 - \omega_2$ is generated by the nonlinear current inside the WSM material. It can be observed both in the transmission geometry, i.e. propagating through the WSM layer, or in the reflection geometry where it propagates away from the WSM surface into the air, opposite to the direction of the incident pump beam. Although there is no incident nonlinear signal, the presence of the ``reflected'' wave is mandated by the boundary conditions, since the nonlinear current exists only on one side of the air-WSM interface.

First, it is instructive to find the linear dispersion and absorption of EM waves propagating in the bulk WSM. Since the material is isotropic within our model, the normal modes are transverse waves with the wave vector magnitude $k=\frac{n(\omega) \omega }{c}$. Here $n(\omega)=\sqrt{\epsilon (\omega)}$ and $\epsilon(\omega) =\epsilon _{b}+\frac{4\pi i\sigma ^{(1)}}{\omega }$, where  $\epsilon _{b}$ is the background 
 dielectric permittivity due to off-resonant transitions to remote bands and $\sigma ^{(1)}$ is the linear response of Weyl fermions given by Eq.~(\ref{sigma1}). The absorption length can be
obtained as $L_{ab}(\omega )= \displaystyle \frac{c}{\omega\mathrm{Im}\left[ n
\right] }$.

\begin{figure}[htb]
\centering
\begin{subfigure}[b]{0.7\textwidth}
\includegraphics[width=1\linewidth]{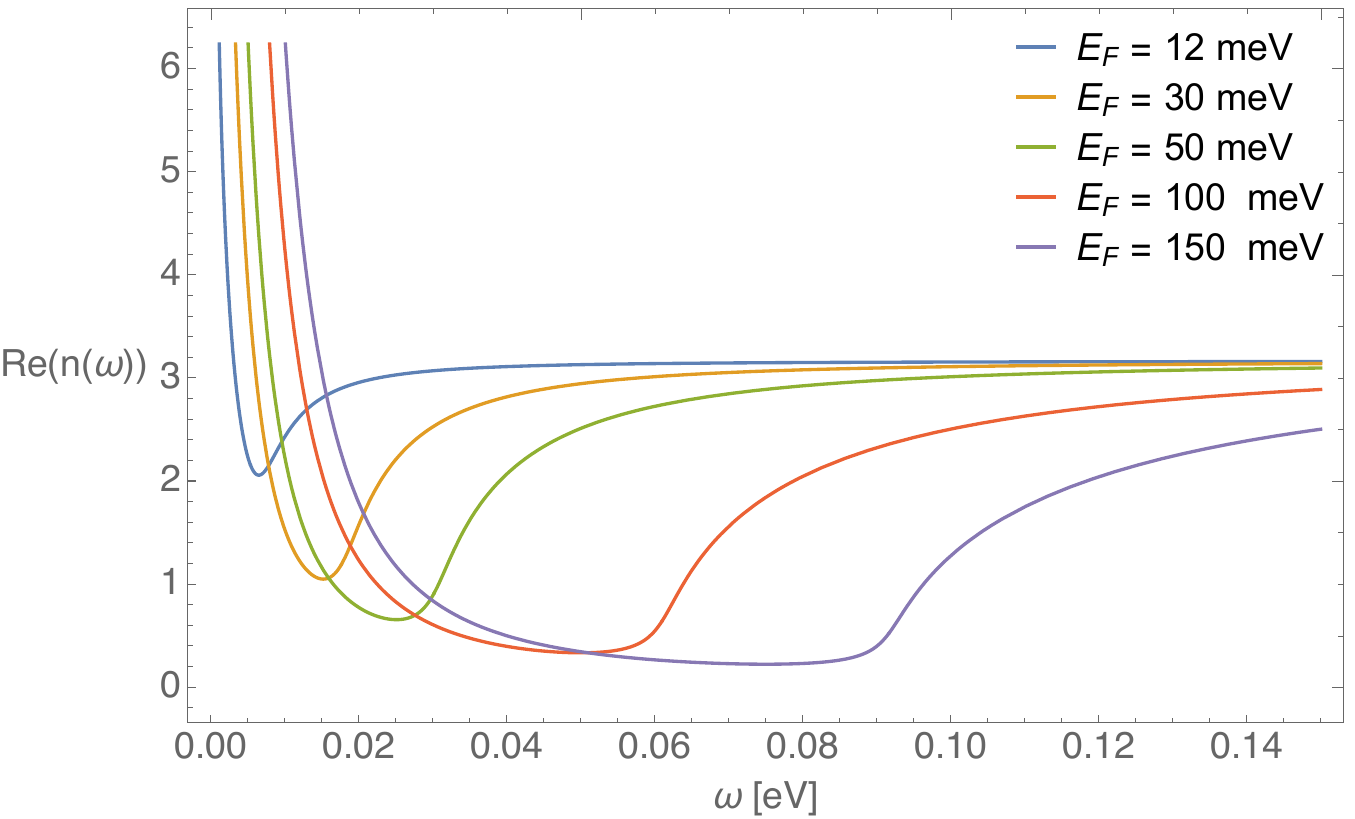}
\caption{ }
\label{fig2a}
\end{subfigure}

\begin{subfigure}[b]{0.7\textwidth}
\includegraphics[width=1\linewidth]{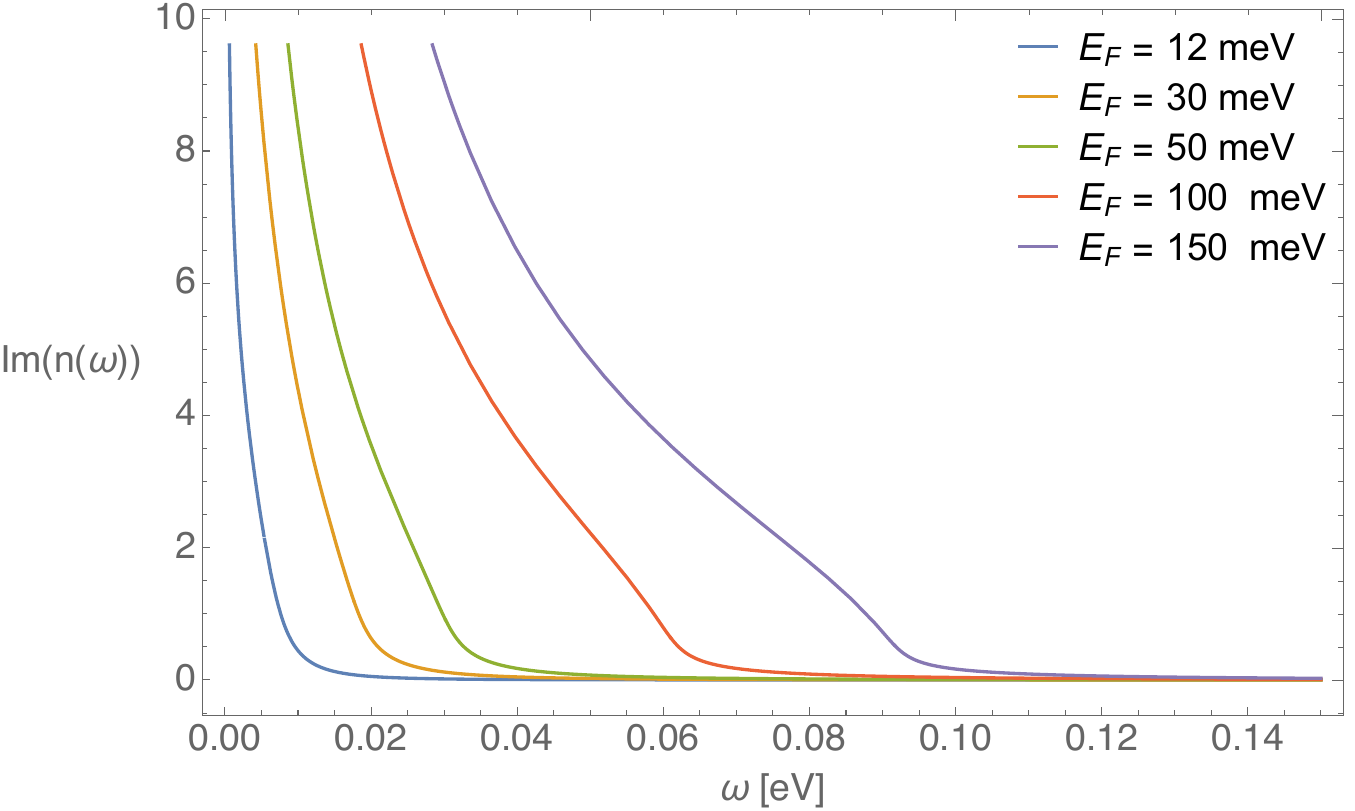}
\caption{  }
\label{fig2b}
\end{subfigure}
\caption{Real (a) and imaginary (b) parts of the linear refractive index as a function of frequency at different Fermi energies for $\epsilon_b = 10$, $\hbar \gamma =  5$ meV, $v_F = 10^8$ cm/s, $g_s = 2$, $g_w = 4$.  }
\end{figure}

Fig.~2 (a,b) show real and imaginary parts of the linear refractive index as a function of frequency at different Fermi energies.  At low frequencies the linear response is dominated by the plasmonic response of Weyl fermions. The plasmonic resonance Re$[\epsilon(\omega)] = 0$ is clearly visible in the refractive index spectra. Below the plasmonic resonance the absorption length drops to the values shorter than the wavelength. Note that the plots cannot be applied to the interband transition region $\hbar \omega > 2 E_F$. 

\begin{figure}[htb]
\begin{center}
\includegraphics[scale=0.5]{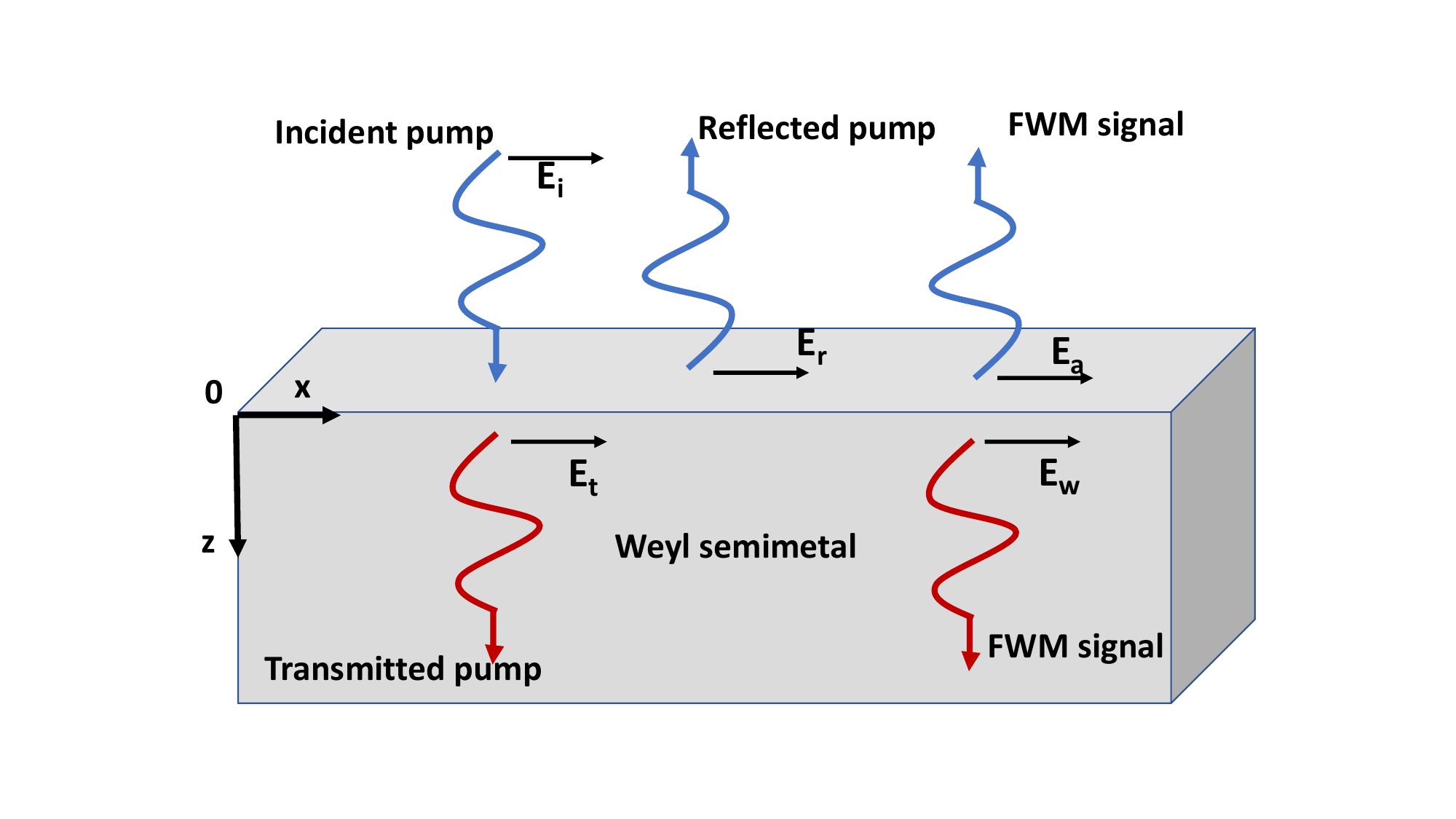}
\caption{Sketch of the simplest experimental geometry. The third-order nonlinear current generated in bulk WSM by incident pump beams gives rise to the FWM signals propagating both into and out of the material. }
\label{sketch}
\end{center}
\end{figure}

Next, we calculate the intensity and power of the nonlinear signal. Assume that the interface between the WSM and the air is in the $(x,y)$ plane and the WSM is at $z > 0$, as shown in Fig.~\ref{sketch}. EM fields in the air above the WSM consist of incident and reflected pump waves, $E_{i1,2} e^{i (\omega_{1,2}/c) z}$ and $E_{r1,2} e^{-i (\omega_{1,2}/c) z}$, and the nonlinear signal wave propagating away from the interface: $E_a = E_a^{(-)} e^{-i k_0 z}$, where $k_0 = \omega_s/c$.  Here we assume that all fields are linearly polarized in the same direction and drop the polarization vectors.

The EM fields in the WSM consist of transmitted pump waves $t_{1,2} E_{i1,2} e^{i k_{1,2} z}$ where $t_{1,2}$ are Fresnel transmission coefficients for the field at frequencies $\omega_{1,2}$, and the co-propagating nonlinear signal which satisfies the wave equation with appropriate boundary conditions at $z = 0$ and the nonlinear polarization $P^{(3)} (\omega_s)$ as the source:
\begin{equation}
\frac{d^2E_w}{dz^2} +  \epsilon(\omega_s) k_0^2 E_w = - \frac{4\pi \omega _{s}^{2}}{
c^{2}}P^{(3)}(\omega _{s}) = A e^{ikz},
\label{wave2}
\end{equation}
where $k \equiv 2 k_1 - k_2$ and 
\begin{equation}
A =- \frac{4\pi \omega _{s}^{2}}{c^{2}} \chi^{(3)} t_1^2 t_2^* E_{i1}^2 E_{i2}^*.  
\label{a}
\end{equation}
Note that the dielectric function $\epsilon(\omega)$ is complex at all frequencies and therefore all relevant wavenumbers are complex: $k_{1,2} = (\omega_{1,2}/c) n_{1,2}$, $k_s = k_0 n_{s}$, where $n_{1,2} = \sqrt{\epsilon(\omega_{1,2})}$, $n_{s} = \sqrt{\epsilon(\omega_{s})}$, and all imaginary parts Im$[n_{1,2,s}]$ are greater than zero. 

The solution to Eq.~(\ref{wave2}) can be written as a sum of the general solution to the homogeneous part and a particular solution to the inhomogeneous equation: 
\begin{equation}
\label{sol}
E_w = E_w^{(+)} e^{ik_s z} + \displaystyle \frac{A}{k_s^2 - k^2} e^{i k z},
\end{equation}
where we dropped the $E_w^{(-)} e^{- ik_s z}$ term. 

The continuity of the tangential electric and magnetic fields at the interface $z = 0$ give $E_a = E_w$ and $\frac{dE_a}{dz} = \frac{dE_w}{dz}$, or 
\begin{equation}
\label{gran}
\begin{array}{rll}
E_a^{(-)} &=& E_w^{(+)}  + \displaystyle \frac{A}{k_s^2 - k^2}  \\
- k_0 E_a^{(-)} &=& k_s E_w^{(+)}  + \displaystyle k \frac{A}{k_s^2 - k^2}. 
\end{array}
\end{equation}
This leads to the following expressions for the nonlinear signal fields propagating from the interface into the air and into the WSM:
\begin{equation}
\label{field}
\begin{array}{rll}
E_a &=& \displaystyle \frac{1}{k_s+ k_0}   \frac{A}{k_s + k} e^{-i k_0 z},  \\
E_w &=& \displaystyle \frac{A}{k_s^2 - k^2} \left( e^{ik z}  -  \frac{k_0 + k}{k_0 + k_s} e^{i k_s z} \right),
\end{array}
\end{equation}
where as a reminder $k = 2 k_1 - k_2$. These expressions can be used to calculate  the nonlinear signal power in both transmission and reflection geometry.  

In the absence of any dissipation (i.e.~when all wavenumbers are real) and for exact phase matching $k_s \rightarrow k$, the monochromatic signal field propagating into the WSM grows linearly with $z$, as expected:
\begin{equation} 
E_w = A \frac{e^{i k_s z}}{2 k_s} \left( \frac{1}{k_0 + k_s} -i z\right); \; E_a = \displaystyle  \frac{A}{2 k_s (k_s + k_0)} e^{-i k_0 z}.
\end{equation} 
Of course for realistic fields of finite duration the region of linear growth of the field is limited by the pulse duration. Moreover, field dissipation is always important because of a fast electron scattering rate $\gamma$ expected in real materials and especially in the region around plasma resonance.  

\begin{figure}[htb]
\begin{center}
\includegraphics[scale=0.5]{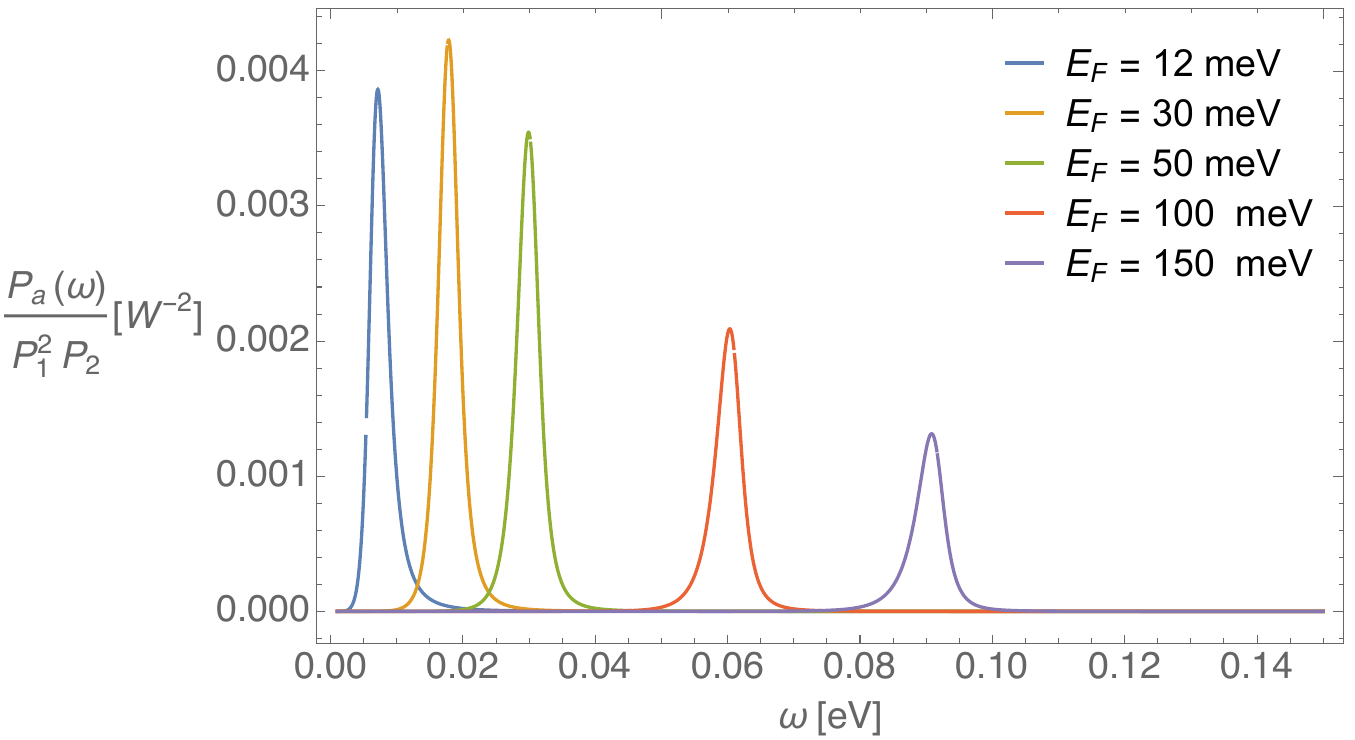}
\caption{The nonlinear signal power in reflection geometry, i.e.~when the signal propagates away from the interface into the air, as a function of frequency and for several values of the Fermi energy.  }
\label{fig4}
\end{center}
\end{figure}

Figure \ref{fig4} shows the power of the nonlinear signal in the reflection geometry, in W per W$^3$ of incident pump power, for degenerate FWM with $\omega_1 = \omega_2$ and assuming that all beams are focused into an area equal to vacuum wavelength squared, i.e. $P_a = \frac{c}{2\pi} |E_a|^2 \left(\frac{2 \pi}{k_0}\right)^2$ and similarly for the pump. All other parameters are the same as in Figs.~1 and 2. 

\begin{figure}[htb]
\begin{center}
\includegraphics[scale=0.5]{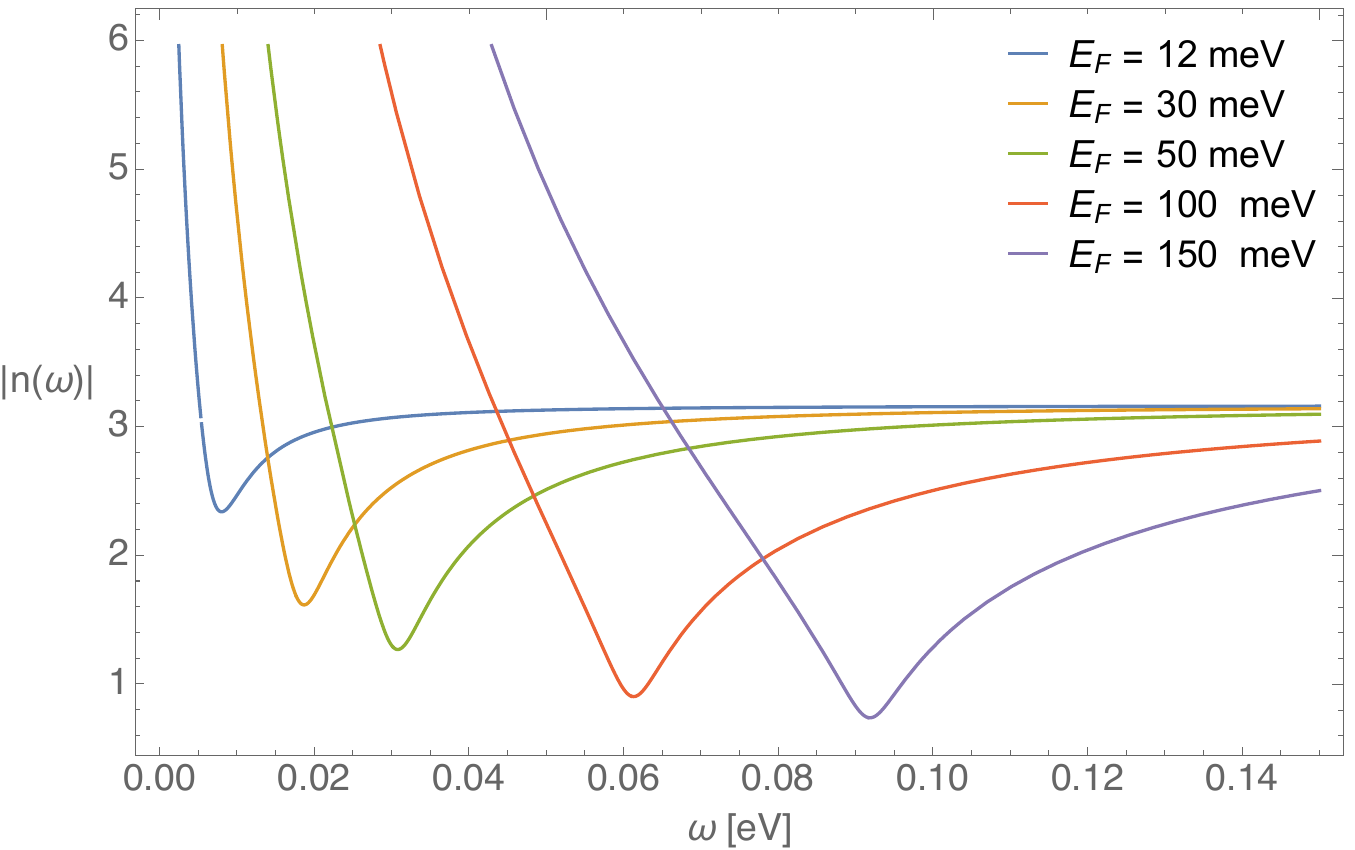}
\caption{Absolute value of the linear refractive index as a function of frequency at different Fermi energies for $\epsilon_b = 10$, $\hbar \gamma =  5$ meV, $v_F = 10^8$ cm/s, $g_s = 2$, $g_w = 4$.  }
\label{fig5}
\end{center}
\end{figure}

The sharp peaks in the spectrum are entirely due to a strong dependence of the signal field intensity from the refractive index of the WSM:
\begin{equation}
\label{ref}
|E_a|^2 = \frac{256 \pi^2 |\chi^{(3)}|^2}{|n_s|^2 |n_s + 1|^8} |E_i|^6.
\end{equation} 
Indeed, the absolute value of the refractive index has a sharp minimum in the vicinity of plasma resonance, see Fig.~\ref{fig5}, which is manifested in the power spectra. Note a simple ``universal '' character of the expression Eq.~(\ref{ref}) for the nonlinear signal, especially given the fact that the value of $\chi^{(3)}$ in this expression does not depend on the Fermi energy. The Fermi energy dependence in Eq.~(\ref{ref}) which is shown in Fig.~\ref{fig4} enters entirely through the refractive index $n_s$. 

The efficiency of the FWM process is quite high, a few mW per W$^3$ of incident pump power, especially in view of the fact that the ``reflected'' nonlinear signal is generated in the subwavelength skin layer below the air/WSM interface. It originates from the high magnitude of $|\chi^{(3)}|$ and strong refractive index dependence mandated by the boundary conditions. The sharp increase in the FWM signal near plasma resonance is conceptually similar to the predicted and observed enhancement of the third-order nonlinear effects for intense laser field propagating in epsilon-near-zero materials; see e.g.~\cite{capretti2015,alam2016} or the recent reviews \cite{reshef2019,kinsley2019} and references therein. 

With detuning from resonance $\delta \omega = \omega_2 - \omega_1 = 0$, the FWM power will decrease following $|\chi^{(3)}|^2 \propto 1/(\delta \omega)^2$ as one can see from Eq.~(\ref{sigma3}) and Fig.~\ref{fig1}. 

The field intensity of the transmitted nonlinear signal in the degenerate FWM process at the distance $z$ into the sample is given by
 \begin{equation}
\label{trans}
|E_w|^2 = \frac{256 \pi^2 |\chi^{(3)}|^2 |E_i|^6}{|n_s|^2 |n_s + 1|^6} \left| \frac{1}{1+n_s} - i k_0 z \right|^2 e^{-2 k_0 {\rm Im}[n_s] z}.
\end{equation} 

\begin{figure}[htb]
\begin{center}
\includegraphics[scale=0.5]{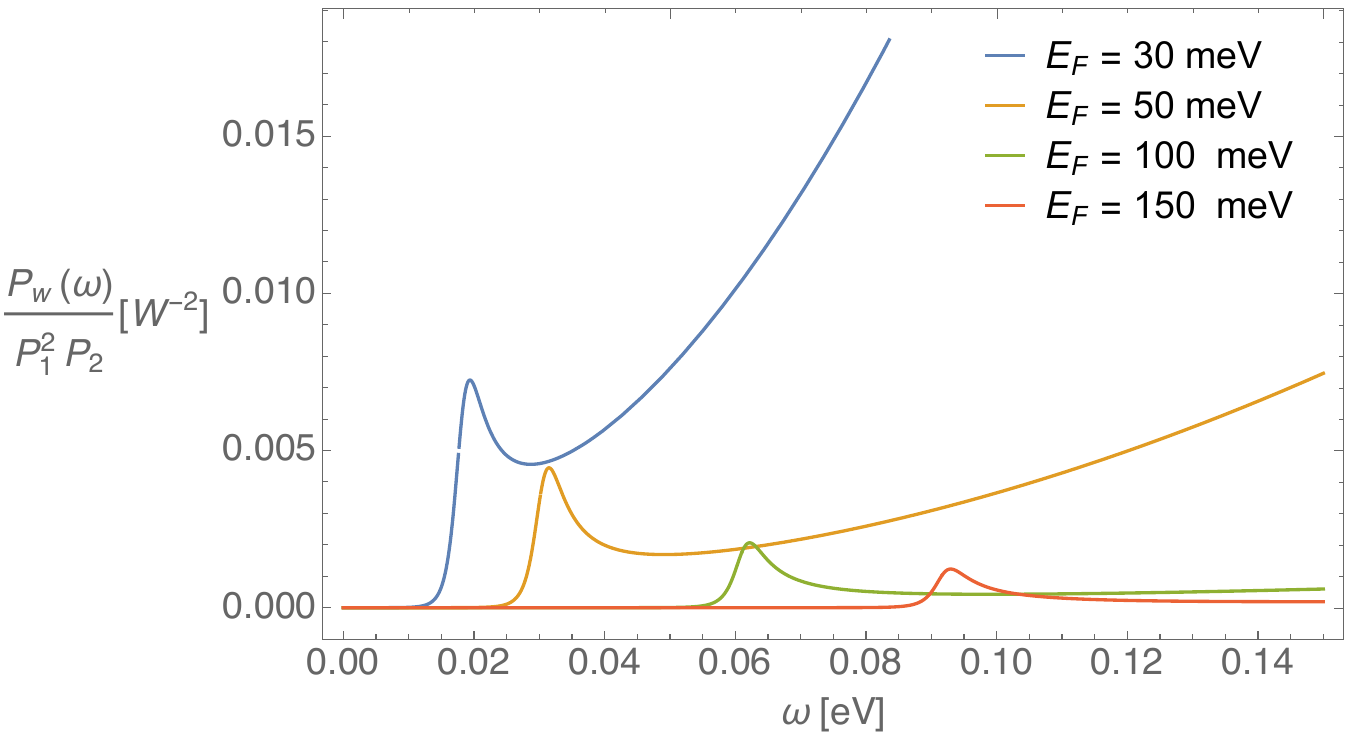}
\caption{The nonlinear signal power after propagating a distance equal to one absorption length $L_{ab} = 1/$Im$[k_s]$ into the sample, as a function of frequency and for several values of the Fermi energy.  }
\label{fig6}
\end{center}
\end{figure}

The corresponding power after propagating the distance equal to the absorption length $L_{ab} = 1/$Im$[k_s]$ into the sample is plotted in Fig.~\ref{fig6} as a function of frequency for different Fermi energies. Here we again assumed that the pump beam was focused into the area of $(2 \pi/k_0)^2$. 

\begin{figure}[htb]
\begin{center}
\includegraphics[scale=0.5]{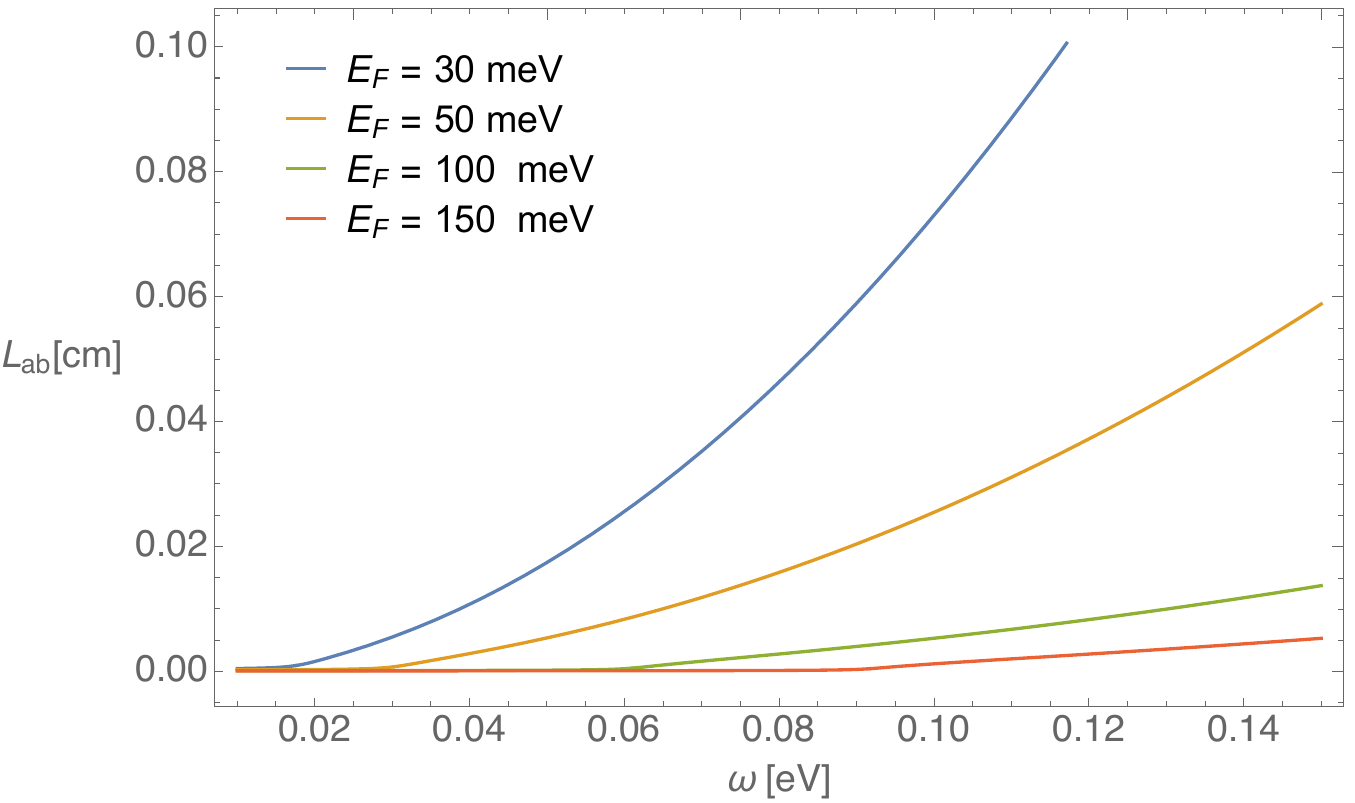}
\caption{Absorption length $L_{ab} = 1/$Im$[k_s]$ as a function of frequency at different Fermi energies for $\epsilon_b = 10$, $\hbar \gamma =  5$ meV, $v_F = 10^8$ cm/s, $g_s = 2$, $g_w = 4$.  }
\label{fig7}
\end{center}
\end{figure}

The characteristic feature of each spectrum is a sharp peak just above plasma resonance, when the refractive index $n_s(\omega)$ is still close to its minimum value, followed by a gradual increase. The gradual increase is entirely due to the absorption length increasing with frequency, as shown in Fig.~\ref{fig7}. Note however that the plots in Figs.~\ref{fig6} and \ref{fig7} cannot be extended beyond $\omega = 2E_F$ where the interband transitions become important. 

\begin{figure}[htb]
\begin{center}
\includegraphics[scale=0.5]{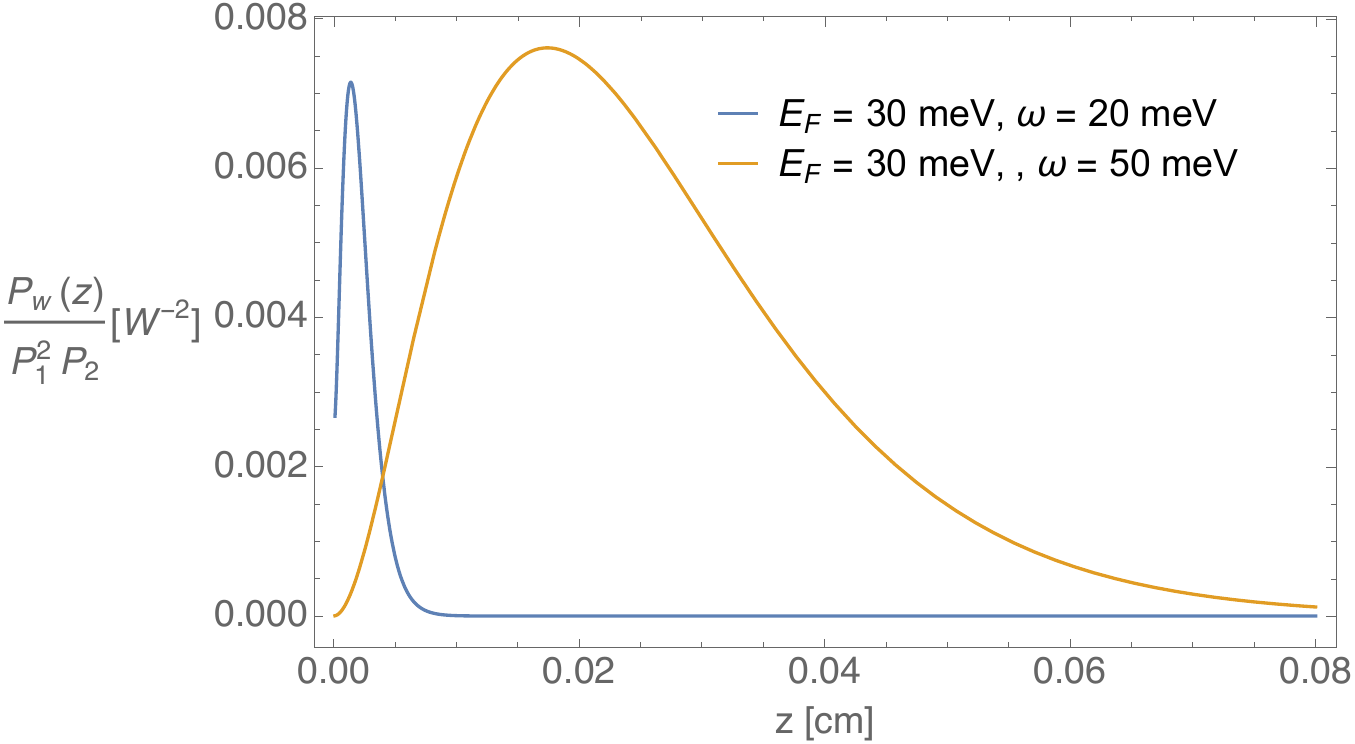}
\caption{The nonlinear signal power as a function of distance $z$ into the sample at two different frequencies and the same Fermi energy.  }
\label{fig8}
\end{center}
\end{figure}

Therefore, for a sample with a given electron density one can get similar levels of the transmitted nonlinear signal power when using a very thin film at frequencies near the plasma resonance and when using thicker films at higher frequencies near the interband transitions cutoff. This is illustrated in Fig.~\ref{fig8} which shows the nonlinear signal power as a function of distance into the sample at two different frequencies and the same Fermi level. 

Various strategies can be employed in order to extract the transmitted nonlinear signal from the sample on the bottom side of the WSM film: an index-matching substrate, tailoring the layer thicknesses to form a Fabry-Perot cavity or a coupled cavity, etc. We won't go into these technical details here. Moreover, since the magnitudes of the signal power in transmission and reflection geometries are similar (compare Fig.~\ref{fig4} and \ref{fig8}), in many cases it is more convenient to use the ``reflected'' (backward-propagating) FWM signal $|E_a|^2$ which is formed in the subwavelength layer of the order of skin depth at the surface.  Then the details of the substrate and actual sample thickness don't matter.

\section{Conclusions}

We studied the nonlinear optical response of Weyl semimetals within the kinetic equation approach which is valid at low enough frequencies in the vicinity of Weyl nodes and below the onset of interband transitions. We calculated the intensity of the nonlinear four-wave mixing signal in both transmission and reflection geometry. The doped bulk WSM exhibits extremely high third order nonlinearity combined with very high absorption loss. This led us to rethink the optimal strategies for nonlinear signal generation. The nonlinear signal intensity is maximized in the vicinity of bulk plasma resonance, which allows one to use ultrathin WSM films of the order of skin depth. The nonlinear generation efficiency turns out to be quite high for a thin film of a highly dissipative material: of the order of  several mW per W$^3$, in both transmission and reflection geometries.  This could pave the way to interesting optoelectronic applications.

\begin{acknowledgments}

This work has been supported in part by the Air Force Office for Scientific Research
through Grant No.~FA9550-17-1-0341 and by NSF Award No.~1936276. M.T. acknowledges the support from RFBR Grant No. 18-29-19091mk and the Federal Research Center Institute of Applied Physics of the Russian Academy of Sciences (Project No. 0035-2019-004).

\end{acknowledgments}

\appendix

\section{Nonlinear optical conductivity derivation}

To evaluate Eq.~(\ref{third}) from the main text for the third-order perturbation of the distribution function we need to calculate  
$G^{(n_{3},m_{3})}G^{(n_{2},m_{2})}g_{1}^{(n_{1},m_{1})}\frac{\partial F}{%
\partial p}$. First, by acting with $G^{(n_{2},m_{2})} \equiv G^{2}$ on $
g_{1}^{(n_{1},m_{1})}\frac{\partial F}{\partial p}=g_{1}^{1}\frac{\partial F%
}{\partial p},$ we obtain
\begin{equation}
\label{one}
G^{2}g_{1}^{1}\frac{\partial F}{\partial p}=g_{1}^{2}g_{1}^{1}\partial
_{p^{2}}^{2}F+g_{2}^{2}\partial _{\phi }g_{1}^{1}\frac{\partial _{p}F}{p}%
+g_{3}^{2}\partial _{\theta }g_{1}^{1}\frac{\partial _{p}F}{p}.
\end{equation}%
Second, acting with $G^{(n_{3},m_{3})}=G^{3}$ on Eq.~(\ref{one}), we get
\begin{eqnarray}
G^{3}G^{2}g_{1}^{1}\frac{\partial F}{\partial p} &=& \left( g_{1}^{3}\partial
_{p}+g_{2}^{3}\frac{\partial _{\phi }}{p}+g_{3}^{3}\frac{\partial _{\theta }%
}{p} \right)  \notag \\
&&\times \left(g_{1}^{2}g_{1}^{1}\partial _{p^{2}}^{2}F+\sin ^{2}\theta
g_{2}^{2}g_{2}^{1}\frac{\partial _{p}F}{p}   + g_{3}^{2}g_{3}^{1}\frac{\partial _{p}F}{p} \right).
\end{eqnarray}

The nonlinear current $\mathbf{j}^{(\omega _{n}=\omega
_{1}+\omega _{2}+\omega _{3})}$ is then given by
\begin{eqnarray}
\left(
\begin{array}{c}
j_{x}^{\omega _{n}} \\
j_{y}^{\omega _{n}} \\
j_{z}^{\omega _{n}}%
\end{array}%
\right) &=&-ev_{F}\int_{0}^{\infty }\int_{0}^{2\pi }\int_{0}^{\pi }\xi
_{n}^{\left( 3\right) }\left(
\begin{array}{c}
\cos \phi \sin \theta \\
\sin \phi \sin \theta \\
\cos \theta%
\end{array}%
\right) p^{2}\sin \theta d\theta d\varphi dp  \notag \\
&=&\frac{1}{3!}\left( I_{i}^{3,2,1}+\, {\rm Permutation}(\omega _{1},\omega
_{2},\omega _{3})\right) ,
\label{curr}
\end{eqnarray}%
where
\begin{eqnarray}
I_{i}^{3,2,1} &=&-ev_{F}\int_{0}^{\infty }\int_{0}^{2\pi }\int_{0}^{\pi
}G^{3}G^{2}g_{1}^{1}\frac{\partial F}{\partial p} \left(
\begin{array}{c}
\cos \phi \sin \theta \\
\sin \phi \sin \theta \\
\cos \theta%
\end{array}%
\right) p^{2}\sin \theta d\theta d\phi dp  \notag \\
&=&\frac{8\pi e^{4}v_{F}F(0)}{15(\gamma -i\omega _{1})(\gamma -i(\omega
_{1}+\omega _{2}))}   \frac{\Delta _{ijkl}E_{1}^{j}E_{2}^{k}E_{3}^{l}}{(\gamma -i(\omega
_{1}+\omega _{2}+\omega _{3}))}
\end{eqnarray}%
where $\Delta _{ijkl}=\delta _{ij}\delta _{kl}+\delta _{ik}\delta
_{jl}+\delta _{il}\delta _{jk}$ and $\delta _{ij}$ is the Kronecker delta. 
Here
we used the relations $\int_{0}^{\infty }\frac{\partial F(p)}{%
p^{2}\partial p}p^{2}dp=-\int_{0}^{\infty }\frac{\partial ^{2}F(p)}{%
p\partial p^{2}}p^{2}dp=-F(0)$ and $\int_{0}^{\infty }\frac{\partial ^{3}F(p)%
}{\partial p^{3}}p^{2}dp=-\int_{0}^{\infty }\frac{\partial \left( \frac{%
\partial F(p)}{p\partial p}\right) }{\partial p}p^{2}dp=-2F(0)$. The summation over repeating indices is assumed.

For a strongly Fermi-degenerate distribution we can replace the equilibrium distribution function with its zero-temperature limit, $F(p)=F(0)\Theta \left( p_{F}-p\right) $, where $F(0)=\frac{%
g_{s}g_{w}}{(2\pi \hbar )^{3}}$. Here $g_{s}$ and $g_{w}$ are the spin and
Weyl node degeneracy, respectively. In this case,  Eq.~(\ref{curr}) becomes
\begin{equation}
j_{i}(\omega _{n}) =\sigma _{ijkl}E_{1}^{j}E_{2}^{k}E_{3}^{l},
\end{equation}
where 
\begin{eqnarray}
&&\sigma _{ijkl}  
= \frac{1}{3!}(\frac{8\pi e^{4}v_{F}g_{s}g_{w}\Delta _{ijkl}}{15(2\pi \hbar
)^{3}(\gamma -i\omega _{1})} 
\frac{1}{\left[ \gamma -i(\omega _{1}+\omega _{2})\right] \left[ \gamma
-i(\omega _{1}+\omega _{2}+\omega _{3})\right] }  \notag \\
&&+\, {\rm Permutation}(\omega _{1},\omega _{2},\omega _{3}))
\end{eqnarray}%
is the third-order nonlinear optical conductivity at zero
temperature.



\begin{thebibliography}{}

\bibitem{Wan2011} X. Wan, A. M. Turner, A. Vishwanath, and S. Y. Savrasov,
Phys. Rev. B 83, 205101 (2011).

\bibitem{Burkov2011} A. A. Burkov and L. Balents, Phys. Rev. Lett. 107, 127205
(2011).

\bibitem{Xu2015} S.-Y. Xu, I. Belopolski, N. Alidoust, M. Neupane, G. Bian, C. Zhang, R. Sankar, G. Chang, Z. Yuan, C.-C. Lee, S.-M. Huang, H. Zheng, J. Ma, D. S. Sanchez, B. Wang, A. Bansil, F. Chou, P. P. Shibayev, H. Lin, S. Jia, and M. Z. Hasan, Science 349, 613 (2015).

\bibitem{Lv2015} B. Q. Lv, H. M. Weng, B. B. Fu, X. P. Wang, H. Miao, et al., Phys. Rev. X 5, 031013 (2015).



\bibitem{Yan2017} B. Yan and C. Felser, Annu. Rev. Condens. Matter Phys. 8,
337 (2017).

\bibitem{Hasan2017} M. Z. Hasan, S.-Y. Xu, I. Belopolski, and S.-M. Huang,
Annu. Rev. Condens. Matter Phys. 8, 289 (2017).

\bibitem{Armitage2018} N. P. Armitage, E. J. Mele, and A. Vishwanath, Rev.
Mod. Phys. 90, 015001 (2018).

\bibitem{Burkov2018} A. A. Burkov, Annu. Rev. Condens. Matter Phys. 9, 359
(2018).

\bibitem{kargarian2015} M. Kargarian, M. Randeria, and N. Trivedi, Sci. Rep. 5, 12683 (2015).

\bibitem{tabert2016-2} C. J. Tabert and J. P. Carbotte, Phys. Rev. B 93, 085442 (2016). 


\bibitem{hofmann2016} J. Hofmann and S. Das Sarma, Phys. Rev. B 91, 241108 (2015).

\bibitem{ukhtary2017} M. S. Ukhtary, A. R. T. Nugraha, and R. Saito, J. Phys. Soc.
Jpn. 86, 104703 (2017). 

\bibitem{felser2017} S. Kimura, H. Yokoyama, H. Watanabe, J. Sichelschmidt,
V. Sus, M. Schmidt, and C. Felser, Phys. Rev. B 96, 075119 (2017). 

\bibitem{kotov2018}  O. V. Kotov and Yu. E. Lozovik, Phys. Rev. B 98, 195446 (2018).

\bibitem{andolina2018} G. M. Andolina, F. M. D. Pellegrino, F. H. L. Koppens, and M.
Polini, Phys. Rev. B 97, 125431 (2018).

\bibitem{zyuzin2018} K. Halterman, M. Alidoust, and A. Zyuzin, Phys. Rev. B 98, 085109 (2018).

\bibitem{rostami2018} H. Rostami and M. Polini, Phys. Rev. B 97, 195151 (2018). 

\bibitem{chen2019} Q. Chen, A. R. Kutayiah, I. Oladyshkin, M. Tokman, and A. Belyanin, Phys. Rev. B 99, 075137 (2019).


\bibitem{narang2019} C. A. C. Garcia, J. Coulter, and P. Narang, https://arxiv.org/abs/1907.04348. 

\bibitem{ma2019} J. Ma, Q. Gu, Y. Liu, J. Lai, P. Yu, X. Zhuo, Z. Liu, J.-H. Chen, J. Feng, and D. Sun, Nat. Mater.  18, 476 (2019). 

\bibitem{moore2019} J. Moore, Nat. Sci. Rev. 6, 206 (2019).  

\bibitem{chen2019-2} Q. Chen, M. Erukhimova, M. Tokman, and A. Belyanin, Phys. Rev. B 100, 235451 (2019). 

\bibitem{tokman2020} I. D.Tokman, Q. Chen, I. A. Shereshevsky, V. I. Pozdnyakova,
I. Oladyshkin, M. Tokman, and A.Belyanin, Phys. Rev. B submitted; https://arxiv.org/pdf/2003.07437.pdf.


\bibitem{okugawa2014} R. Okugawa and S. Murakami, Phys. Rev. B 89, 235315 (2014).

\bibitem{capretti2015} A. Capretti, Y. Wang, N. Engheta, and L. D. Negro, Opt. Lett. 40, 1500 (2015). 

\bibitem{alam2016} M. Z. Alam, I.D. Leon, R. W. Boyd, Science 352, 795 (2016). 

\bibitem{Sun 2018} Z. Sun, D. N. Basov, M. M. Fogler, Phys. Rev. B 97,
075432 (2018)

\bibitem{Zhong 2019} Y. Zhong, W. Feng, Z. Liu, C. Zhang and, J.C. Cao, Physica B: Condensed Matter 555,
81--84 (2019).

\bibitem{zuber2020} J. W. Zuber, T. Zhao, S. Gong, M. Hu, R. B. Zhong, C. Zhang , and S. G. Liu, Phys. Rev. B 101, 085307 (2020). 

\bibitem{wang2016} Y. Wang, M. Tokman, and A. Belyanin, 
Phys. Rev. B 94, 195442 (2016).

\bibitem{Boyd2003} R.W. Boyd, \textit{Nonlinear Optics} (Academic Press, London, 2003).


\bibitem{reshef2019} O. Reshef, I.D. Leon,  M. Z. Alam, and R. W. Boyd, Nat. Rev. Materials 4, 535 (2019). 

\bibitem{kinsley2019} N. Kinsley and J. Khurgin, Opt. Mat. Express 9, 2793 (2019). 



\end{thebibliography}
\end{document}